\documentclass[journal=mamobx,manuscript=article,layout=twocolumn]{achemso}

\usepackage{graphicx}
\usepackage{dcolumn}
\usepackage{float}
\usepackage{bm}
\usepackage{amssymb}
\usepackage{amsmath}

\title{Free energy of a folded semiflexible polymer confined to a nanochannel of various
geometries}

\author{James M. Polson}
\email{jpolson@upei.ca}
\affiliation{
Department of Physics, University of Prince Edward Island, 550 University Ave.,
Charlottetown, Prince Edward Island, C1A 4P3, Canada
}

\begin{document}


\begin{abstract}
Monte Carlo simulations are used to study the conformational properties
of a folded semiflexible polymer confined to a long channel. We measure the variation in
the conformational free energy with respect to the end-to-end distance of the polymer,
and from these functions we extract the free energy of the hairpin fold, as well
as the entropic force arising from interactions between the portions of the polymer
that overlap along the channel. We consider the scaling of the free energies with respect
to varying the persistence length of the polymer, as well as the channel
dimensions for confinement in cylindrical, rectangular and triangular channels.
We focus on polymer behaviour in both the classic Odijk and backfolded Odijk
regimes. We find the scaling of the entropic force to be close to that predicted from
a scaling argument that treats interactions between deflection segments at the
second virial level. In addition, the measured hairpin fold free energy is consistent
with that obtained directly from a recent theoretical calculation for cylindrical channels.
It is also consistent with values determined from measurements of the global persistence
length of a polymer in the backfolded Odijk regime in recent simulation studies.
\end{abstract}

\maketitle

\section{Introduction}
\label{sec:intro}

Polymers confined to channels of a width smaller than the radius of gyration stretch 
along the channel. In recent years, there has been considerable interest in elucidating 
the conformational behavior arising from such confinement on polymers such as DNA.%
\cite{dai2016polymer,reisner2012dna} A thorough understanding of this behavior is
expected to be of value in the development of a variety of applications that exploit the 
effects of confinement on polymers, including DNA sorting,\cite{dorfman2012beyond} DNA 
denaturation mapping,\cite{reisner2010single,marie2013integrated} and genome 
mapping.\cite{lam2012genome,dorfman2013fluid}

The conformational behavior of a polymer in a channel is determined by the effective 
width, $w$, the persistence length, $P$, and the contour length, $L$ of the polymer, 
as well as the lateral dimension(s) of the channel, $D$. Several 
distinct conformational regimes have been identified, each with 
its own scaling properties for the extension length and its variance, as well as the 
free energy. For sufficiently narrow channels where $P/D\gg 1$, the polymer is 
highly extended and lies in the Odijk regime\cite{odijk1983statistics} with a 
Gaussian distribution of extension lengths for sufficiently large $L$. At 
the opposite extreme of wide channels such that $D\gg P^2/w$ lies the classic 
de~Gennes regime.\cite{deGennes_book} Between these limits lie the extended de-Gennes 
regime\cite{dai2014extended,werner2014confined,smithe2015finite} and the backfolded 
Odijk regime.  The latter regime was predicted by Odijk\cite{odijk2006dna,odijk2008scaling} 
and was later fully characterized using computer simulations.\cite{muralidhar2014backfolding,%
muralidhar2016backfolded,muralidhar2016backfolding} For rectangular channels with two 
independent lateral dimensions, there are additional scaling regimes.\cite{werner2015scaling} 
The scaling theories used in these studies employ concepts such as deflection segments, hairpins 
and blobs to describe local structures that define the separate scaling regimes.  
Recently, a theory has been developed that avoids such distinctions between
microscopic models and accurately describes the statistics of a channel-confined 
polymer in the range $D\lesssim P^2/w$.\cite{werner2017one} The scaling predictions
for the Odijk, backfolded Odijk and extended de~Gennes regimes emerge as limiting cases
of the more general theory.

Polymer folding in nanochannels results in internal overlap of the polymer along the
channel axis. Folding is an essential feature in the backfolded Odijk regime, in which
elongated sections of polymer are connected by hairpin turns. Hairpin formation for 
$D\lesssim P$ has recently been observed in experiments of DNA coated by the protein RecA 
and confined to rectangular channels.\cite{werner2018hairpins} In addition to folding in 
equilibrium states, a polymer can be inserted into a nanochannel in a nonequilibrium folded 
state, whereupon it proceeds to unfold and thereby reduce its internal overlap 
until it reaches its equilibrium extension. Such conformational relaxation has been 
observed in various experimental studies of DNA. For example, Levy {\it et al.} examined 
the unfolding of DNA that was electrophoretically driven into a nanochannel in a folded 
state.\cite{levy2008entropic} Alizadehheidari {\it et al.} studied the unfolding 
of an initially circular DNA molecule that was subjected to a light-induced double-strand 
break.\cite{alizadehheidari2015nanoconfined} They observed an increase in the unfolding
rate with increasing extension induced by decreasing the buffer concentration and
thus increasing the confinement.
More recently, Reifenberger {\it et al.} studied the unfolding dynamics of DNA confined 
in square channels of width $D$=45~nm such that $P\approx D$.\cite{reifenberger2018odijk}
The unfolding rates are consistent with predictions using a deterministic model 
that combined the Odijk excluded volume\cite{odijk2008scaling} with numerical calculations
of the Kirkwood diffusivity for confined DNA.

DNA unfolding is driven by excluded volume interactions between the portions of the 
molecule that overlap along the channel axis. These interactions tend to reduce the 
conformational entropy by an amount that increases as the degree of overlap increases. 
Computer simulations of unfolding of flexible hard-sphere chains in channels have corroborated 
this general picture.\cite{ibanez2012entropic,ibanez2013hairpin} A steeper gradient
in the overlap free energy is expected to increase the unfolding rate. In addition,
the scaling properties of the overlap free energy, as well as the free energy of
hairpin formation, are integral to theories of the backfolded Odijk 
regime.\cite{odijk2006dna,odijk2008scaling,chen2017conformational} 
For this reason, it is of interest to characterize the variation of the conformational 
free energy with the degree of internal polymer overlap. Recently, we employed Monte Carlo 
simulations to measure such free energy functions for polymers confined to cylindrical 
channels.\cite{polson2017free} For the case where $P\gtrsim D$, we found that the 
scaling of the overlap free energy gradient with respect to $D$ and $P$ was quantitatively 
consistent with the predicted scaling. In addition, the hairpin
free energy was appreciably lower than that predicted by Odijk,%
\cite{odijk2006dna,odijk2008scaling} a result also observed by Muralidhar {\it et al.},%
\cite{muralidhar2014backfolding,muralidhar2016backfolded} and was closer
to the value predicted by Chen.\cite{chen2017conformational}

In the present study, we continue the characterization of free energy functions for
polymers in the regime $D\lesssim P$, focusing mainly on the effects of varying the
channel geometry. In addition to cylindrical channels, we now consider the cases 
of rectangular and triangular channels. While rectangular channels are the most
commonly used in DNA extension experiments, triangular channels are also
widely employed.\cite{fanzio2011dna,angeli2011dna,fanzio2012modulating,%
manneschi2013conformations,manneschi2014stretching}
In addition, we consider the effects of varying the asymmetry of the channel cross section, 
which is known to affect the extension behavior and scaling regimes of 
rectangular\cite{werner2015scaling} and triangular\cite{reinhart2013entropic} channels.
Generally, we find the overlap free energy scales in a manner consistent with the
theoretical predictions, with small deviations in the predicted scaling exponents, as
observed previously for cylindrical channels.\cite{polson2017free} The asymmetry in 
the channel shape has negligible effect on the overlap free energy gradient, in contrast 
to an appreciable decrease in the hairpin free energy with increasing asymmetry. 
We also find the measured hairpin free energy in rectangular channels is quantitatively 
consistent with the values obtained from the simulations of Muralidhar {\it et al.}%
\cite{muralidhar2014backfolding,muralidhar2016backfolded} and significantly less
than those predicted by Odijk.\cite{odijk2006dna,odijk2008scaling}
Finally, we examine the fluctuations in the shape and orientation of the hairpin turns 
that underly this discrepancy. For cylindrical channels, the observed behavior is
consistent with that predicted by Chen.\cite{chen2017conformational}

\section{Model}
\label{sec:model}

We employ a minimal model to describe a semiflexible polymer confined to an infinitely long 
channel.  The polymer is modeled as a chain of hard spheres, each with 
diameter $\sigma$.  The pair potential for non-bonded monomers is thus $u_{\rm{nb}}(r)=\infty$ 
for $r\leq\sigma$ and $u_{\rm{nb}}(r)=0$ for $r>\sigma$, where $r$ is the distance between 
the centers of the monomers. Pairs of bonded monomers interact with a potential 
$u_{\rm{b}}(r)= 0$ if $0.9\sigma<r<1.1\sigma$, and $u_{\rm{b}}(r)= \infty$, otherwise.  
The stiffness of the polymer chain is modeled using a bending potential associated with 
each consecutive triplet of monomers. The potential has the form, 
$u_{\rm bend}(\theta) = \kappa (1-\cos\theta)$.  The angle $\theta$ is defined at monomer 
$i$ such that $\cos\theta_i \equiv \hat{u}_{i}\cdot\hat{u}_{i+1}$,
where $\hat{u}_i$ is a normalized bond vector pointing from monomer $i-1$ to monomer $i$.
The bending constant $\kappa$ determines the stiffness of the polymer and is related
to the persistence length $P$ by\cite{micheletti2011polymers}
$\exp(-\langle l_{\rm bond}\rangle/P) = \coth(\kappa/k_{\rm B}T) - k_{\rm B}T/\kappa$. 
For our model, the mean bond length is $\langle l_{\rm bond} \rangle \approx \sigma$. 
%
%

The channel has constant cross-sectional area and is constructed with hard walls,
such that the monomer-wall interaction energy is $u_{\rm w}(r) = 0$ if monomers do not
overlap with the wall and $u_{\rm w}(r) = \infty$ if there is overlap. We consider
walls with cross-sections that are circular (i.e. cylindrical channels), rectangular, 
and triangular. We define the channel cross-sectional area $A$ to be that for the space within
the channel that is accessible to the centers of the monomers. Further, we define
the effective channel width to be $D=\sqrt{4A/\pi}$. Thus, in the case of cylindrical 
channels, the effective channel width is related to its true diameter according
to $D=D_{\rm true}-\sigma$. For rectangular channels, we define the cross-section
aspect ratio $r$ as the ratio of the lateral dimensions of the space in the channel
accessible to the monomer centers. As an illustration, if the true lateral dimensions are 
$7\sigma \times 5\sigma$, then the aspect ratio is $r=(6\sigma/4\sigma)=1.5$ and
$D=\sqrt{4(6\sigma\times 4\sigma)/\pi}=5.53\sigma$. Note that the special case of square
channels corresponds to $r$=1. For triangular channels, we consider only channels
with cross-sections that are isosceles triangles. We denote the angle of the apex
of the triangle as $\theta_{\rm ap}$; thus, the angles of the other two vertices are
each $(\pi-\theta_{\rm ap})/2$.

\section{Methods}
\label{sec:methods}

For the model systems described above, Monte Carlo simulations were used to calculate the 
free energy as a function of the end-to-end distance of the polymer, $\lambda$, as measured 
along the axis of the confining channel.  The simulations employed the Metropolis algorithm and 
the self-consistent histogram (SCH) method.\cite{frenkel2002understanding} The SCH method 
efficiently calculates the equilibrium probability distribution ${\cal P}(\lambda)$, and 
thus its corresponding free energy function, $F(\lambda) = -k_{\rm B}T\ln {\cal P}(\lambda)$. 
We have previously used this procedure to measure free energy functions in our recent
study of polymer folding in cylinders\cite{polson2017free}, as well as in simulation studies
of polymer segregation\cite{polson2014polymer,polson2018segregation} and polymer translocation.%
\cite{polson2013simulation,polson2013polymer,polson2014evaluating,polson2015polymer}

To implement the SCH method, we carry out many independent simulations, each of which employs a
unique ``window potential'' of the form:
\begin{eqnarray}
{W_i(\lambda)}=\begin{cases} \infty, \hspace{8mm} \lambda<\lambda_i^{\rm min} \cr 0,
\hspace{1cm} \lambda_i^{\rm min}<\lambda<\lambda_i^{\rm max} \cr \infty,
\hspace{8mm} \lambda>\lambda_i^{\rm max} \cr
\end{cases}
\label{eq:winPot}
\end{eqnarray}
where $\lambda_i^{\rm min}$ and $\lambda_i^{\rm max}$ are the limits that define the range
of $\lambda$ for the $i$-th window.  Within each window of $\lambda$, a probability
distribution $p_i(\lambda)$ is calculated in the simulation. The window potential width,
$\Delta \lambda \equiv \lambda_i^{\rm max} - \lambda_i^{\rm min}$, is chosen to be
sufficiently small that the variation in $F$ does not exceed 2--3 $k_{\rm B}T$.
The windows are chosen to overlap with half of the adjacent window, such that
$\lambda^{\rm max}_{i} = \lambda^{\rm min}_{i+2}$.  The window width was typically
$\Delta \lambda = 2\sigma$. The SCH algorithm was employed to reconstruct the unbiased
distribution, ${\cal P}(\lambda)$ from the $p_i(\lambda)$ histograms.  For further
detail of the histogram reconstruction algorithm, see Ref.~\citenum{frenkel2002understanding}.

Polymer configurations were generated by carrying out single-monomer moves using a combination of
translational displacements and crankshaft rotations, as well as reptation moves.
Trial moves were accepted with a probability $p_{\rm acc}$=${\rm min}(1,e^{-\Delta E/k_{\rm B}T})$, 
where $\Delta E$ is the energy difference between trial and current states. 
Most simulations employed a polymer of length $N$=200 monomers, which is 
sufficient to obtain reliable values of the free energy gradient in the overlap regime and 
the hairpin free energy.\cite{polson2017free} The system was equilibrated
for typically $10^7$ MC cycles, following which a production run of $4\times10^8$
MC cycles was carried out.  On average, during each MC cycle a displacement or rotation
move for each monomer and a reptation move is attempted once.

In the results presented below, quantities of length are measured in units of $\sigma$ and
energy in units of $k_{\rm B}T$. 

\section{Results}
\label{sec:results}

\begin{figure}[!ht]
\begin{center}
\includegraphics[width=0.43\textwidth]{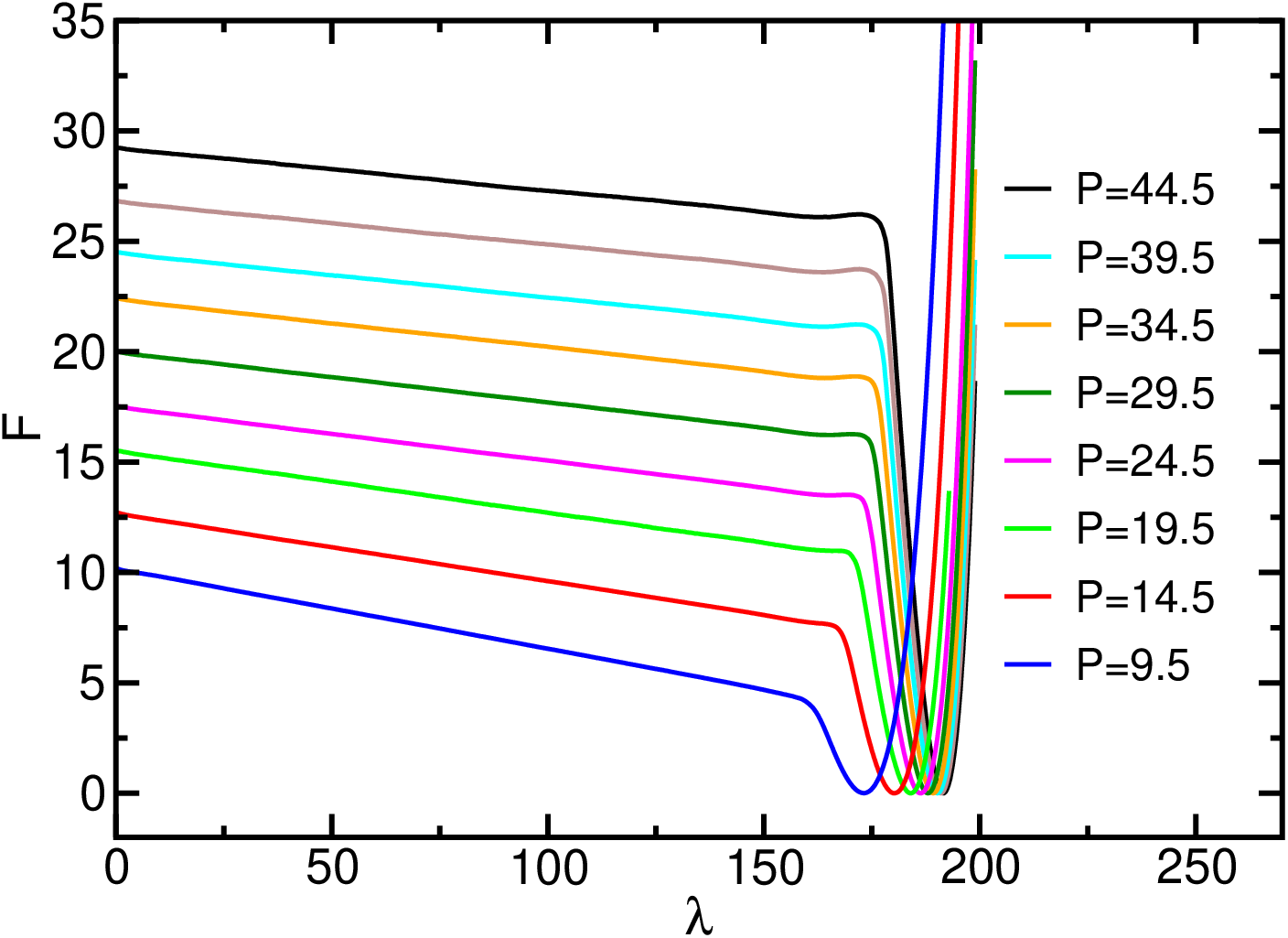}
\end{center}
\caption{Free energy functions for a polymer of length $N$=200 in a square channel
with $D$=6, where the effective diameter is defined $D\equiv \sqrt{4A/\pi}$ for a channel
of effective cross-sectional area $A$, as defined in the text. Results for various 
persistence lengths are shown. The functions are each shifted such that $F$=0 at the minimum.}
\label{fig:F.N200.square.wid6.317}
\end{figure}

Figure~\ref{fig:F.N200.square.wid6.317} shows free energy functions $F(\lambda)$ for
a polymer of length $N$=200 in a square tube of effective diameter $D$=6. 
Results for various persistence lengths are shown. The curves are all qualitatively
similar, and the general trends are also observed for channels of other geometries.
At sufficiently high $\lambda$, there is a free energy well with a depth that 
increases with increasing $P$. The location of the minimum is a measure of the equilibrium
extension length of the polymer, and the width of the well determines the magnitude
of the fluctuations in the extension length. As $P$ increases, the free energy minimum narrows 
and shifts to higher $\lambda$; thus, the extension length increases and the fluctuations 
decrease, consistent with established results for the Odijk regime.\cite{odijk1983statistics} 
As $\lambda$ decreases the free energy initially rises, then abruptly transitions to a 
regime where $F$ increases at a much slower rate with decreasing $\lambda$. 
The well depth, $\Delta F_{\rm w}$, is defined as the difference in the free energy 
between its minimum and the value at the transition location.
Below this transition point, $F$ increases linearly with decreasing $\lambda$ down to
$\lambda$=0. At sufficiently high $P$, an intermediate regime is present in which
$F$ is approximately constant. This regime occurs over a narrow range of $\lambda$
that increases slightly with increasing $P$. In the linear regime, the free energy gradient,
$f\equiv |dF/d\lambda|$, decreases as the polymers become more rigid.

The origins of the qualitative trends observed in Fig.~\ref{fig:F.N200.square.wid6.317} are 
straightforward and were mostly explained in Ref.~\citenum{polson2017free}. The steep rise
in $F$ as $\lambda$ decreases from $\lambda_{\rm min}$ is associated with the formation
of a hairpin turn in the polymer.  At the value of $\lambda$ where $F$ is at the top of 
the well, the polymer has a ``candy-cane'' shape with a fully formed hairpin.
Thus, $\Delta F_{\rm W}$ is a measure of the free energy associated with hairpin formation. 
As the polymer stiffens, the energy required to bend the polymer backwards also increases,
and this contributes to the increase in $\Delta F_{\rm w}$ with $P$. Note that there are
also important contributions of the entropy of the hairpin to $\Delta F_{\rm w}$,%
\cite{odijk2006dna,chen2017conformational} as will be discussed further below.
As $\lambda$ further decreases below the point where the hairpin has formed, 
portions of the polymer begin to overlap along the channel. As $\lambda$ decreases the 
degree of overlap increases, and the excluded volume interactions between these strands 
reduce the conformational entropy and cause $F$ to increase.
The intermediate regime of constant $F$ occurs when one of the two strands 
is sufficiently short that the probability of a collision with the longer strand
is negligible. This arises because the strands on either side of the hairpin
initially track along the channel walls on opposite sides of the channel. Only when both 
strands are sufficiently long will a collision be likely. This avoidance persists
longer for stiffer chains, leading to an increase in the width of the nearly-constant-$F$
regime with increasing $P$.

As noted earlier, it is of interest to investigate the dependence of both the free energy 
gradient in the overlap regime, $f$, and the free energy of the hairpin, $\Delta F_{\rm w}$, 
on the key system parameters. These include the polymer persistence and contour lengths, 
the chain width, $w$, and the channel dimensions. In addition, it is useful to 
compare the trends for channels of different geometries. To make such a comparison 
meaningful, we consider results for equal values of $A$, the cross-sectional area
of the channel accessible to the monomer centers and, thus, for the same
effective channel diameter, $D\equiv \sqrt{4A/\pi}$. 

\begin{figure}[!ht]
\begin{center}
\includegraphics[width=0.46\textwidth]{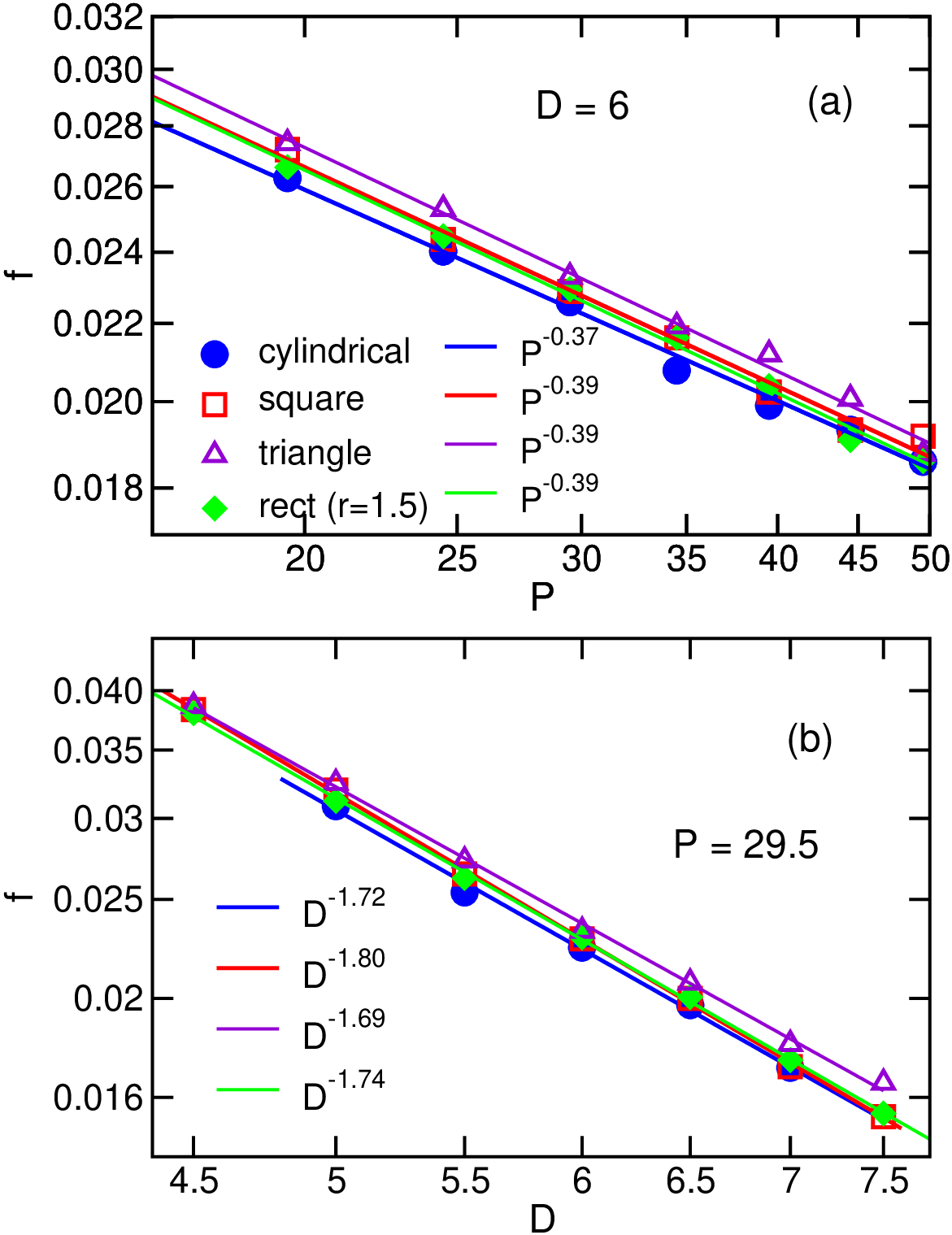}
\end{center}
\caption{(a) Free energy gradient $f\equiv |dF/d\lambda|$ vs $P$ for cylindrical, square,
and (equilateral) triangular channels, as well as for rectangular channels with an aspect ratio 
of $r$=1.5.  Results are shown for a polymer with $N$=200 in channels of effective diameter 
$D$=6.  The solid lines are power law fits to the data sets.  (b) As in (a), except $f$ vs 
$D$ for fixed $P$=29.5. The data for cylindrical channels are taken from 
Ref.~\citenum{polson2017free}.}
\label{fig:dFdk.P.D}
\end{figure}

\begin{figure}[!ht]
\begin{center}
\includegraphics[width=0.43\textwidth]{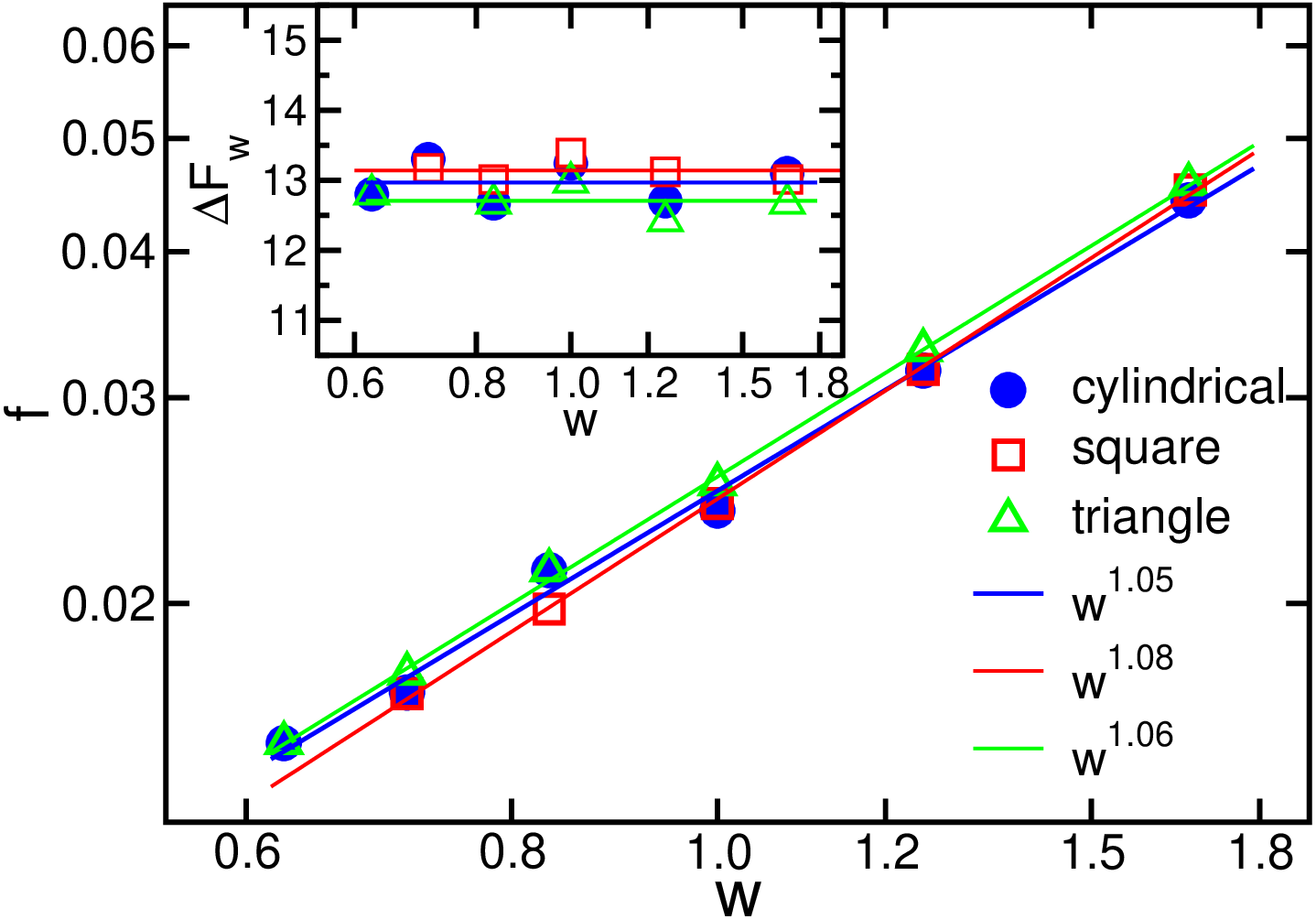}
\end{center}
\caption{Free energy gradient $f\equiv |dF/d\lambda|$ vs the polymer width $w$. Results are 
shown for cylindrical, square and (equilateral) triangular channels for $P/D$=4.08, such that 
$D/w$=6 and $P/w$=24.5 for $w$=1. The solid lines show fits to the power law 
$f\sim w^{\gamma}$.  The inset shows the free energy well depth vs $w$ for the same systems.}
\label{fig:f.width}
\end{figure}

Figure \ref{fig:dFdk.P.D}(a) shows the variation of $f$ with $P$ for fixed $D$=6, while
Fig.~\ref{fig:dFdk.P.D}(b) shows the variation with $D$ for fixed $P$=29.5. In each case,
results are shown for cylindrical, square and triangular channels, as well as a 
rectangular channel with an asymmetry factor of $r$=1.5. The free
energy gradient scales with persistence length as $f\sim P^{-\alpha}$, where $\alpha$
lies in the range $\alpha$=0.37--0.39 for the different channel geometries. Likewise,
the results in Fig.~\ref{fig:dFdk.P.D}(b) show that $f\sim D^{-\beta}$, where 
$\beta$=1.69--1.80 for the different geometries. Figure~\ref{fig:f.width} shows the
variation of $f$ with the polymer width, $w$ at fixed $P/D$=4.08, where $w$ is defined
such that $D/w$=6 and $P/w$=24.5 for $w$=1. The scaling is approximately linear, such
that $f\sim w^\gamma$, where $\gamma$=1.05--1.08 for the different geometries.
Note that the results for the scaling with $P$ and $D$ for cylindrical channels were
reported earlier.\cite{polson2017free}

The observed scaling of the free energy gradient with $D$, $P$ and $w$ can be understood using
the model originally employed by Odijk to describe the backfolding regimes of DNA confined to a
channel.\cite{odijk2008scaling} This approach was also used recently to analyze folding in
simulations of semiflexible polymers in cylindrical channels.\cite{polson2017free}
The expression for the free energy gradient in the folding regime is derived as follows.
The overlapping polymer strands are modeled as rigid rods with a length equal to the Odijk 
deflection length $l_{\rm d}\sim (D^2P)^{1/3}$. At the 2nd virial level,
the interaction free energy is given by by $F^{\rm (int)}/k_{\rm B}T = (l_{\rm d}^2w N^2/V)
\langle |\sin\delta|\rangle$ for $N$ rods of width $w$ confined to a volume $V$, where 
$\delta$ is the angle between the rods.\cite{odijk1986theory} When the rods are highly aligned, 
$\langle|\sin\delta|\rangle \sim \sqrt{\langle\theta^2\rangle}$, where $\theta$ is the angle 
between the rod and the alignment direction.  Assuming that the alignment arises principally 
from confinement, it follows $\langle\theta^2\rangle \sim (D/l_{\rm d})^2$.  Further, we note
that $V$ is the volume over which the intermolecular segments overlap. This is given
by $V\sim l_{\rm ov} A$, where $A$ is the cross-sectional area and the overlap region length 
is $l_{\rm ov} \approx (L-h-\lambda)/2$, where $L$ is the contour length of the polymer
and $h$ is the length of the hairpin turn. Using the effective dimension of the channel 
$D$ such that $A\propto D^2$, and noting that the number of deflection lengths of the two 
overlapping strands is $N=2l_{\rm ov}/l_{\rm d}$, it can be shown that
$F\sim (L-h-\lambda)w D^{-5/3}P^{-1/3}$ plus terms independent of $\lambda$.
Consequently, the free energy gradient is predicted to scale as
\begin{eqnarray}
f \sim w D^{-5/3}P^{-1/3}.
\label{eq:fscale}
\end{eqnarray}
Note that this model predicts that the scaling properties of $f$ are independent of channel 
geometry.  The predicted scaling is close to that observed
for each geometry examined.  As noted previously,\cite{polson2017free} the discrepancy may 
be due to the fact that the system only marginally satisfies the condition for the Odijk 
regime that $P\gg D$, the assumption that the deflection segments are sufficiently 
aligned, modeling the deflection segments as rigid rods, as well as the inadequacy of the
2nd virial approximation for modeling interactions between deflection segments. 

Note that the magnitude of the free energy gradient in Fig.~\ref{fig:dFdk.P.D} tends 
to be slightly greater for triangular channels compared to that for square or cylindrical 
channels. This likely arises from entropic depletion near the corners of the triangle, 
an effect that has been noted previously.\cite{manneschi2013conformations,%
reinhart2013entropic} This depletion reduces the effective cross-section area, leading to 
an increase in the likelihood of collisions between overlapping portions of the polymer and 
thus a greater rate of increase in the free energy with overlap. In 
Ref.~\citenum{manneschi2013conformations} it was observed that the extension length 
for DNA in triangular channels was greater than that for square channels of the same 
cross-section. The increase in extension also arises from the same reduction in the
effective area. The magnitude of that increase was very small, consistent with the small 
increase in $f$ observed for triangular channels in the present work.

\begin{figure}[!ht]
\begin{center}
\includegraphics[width=0.43\textwidth]{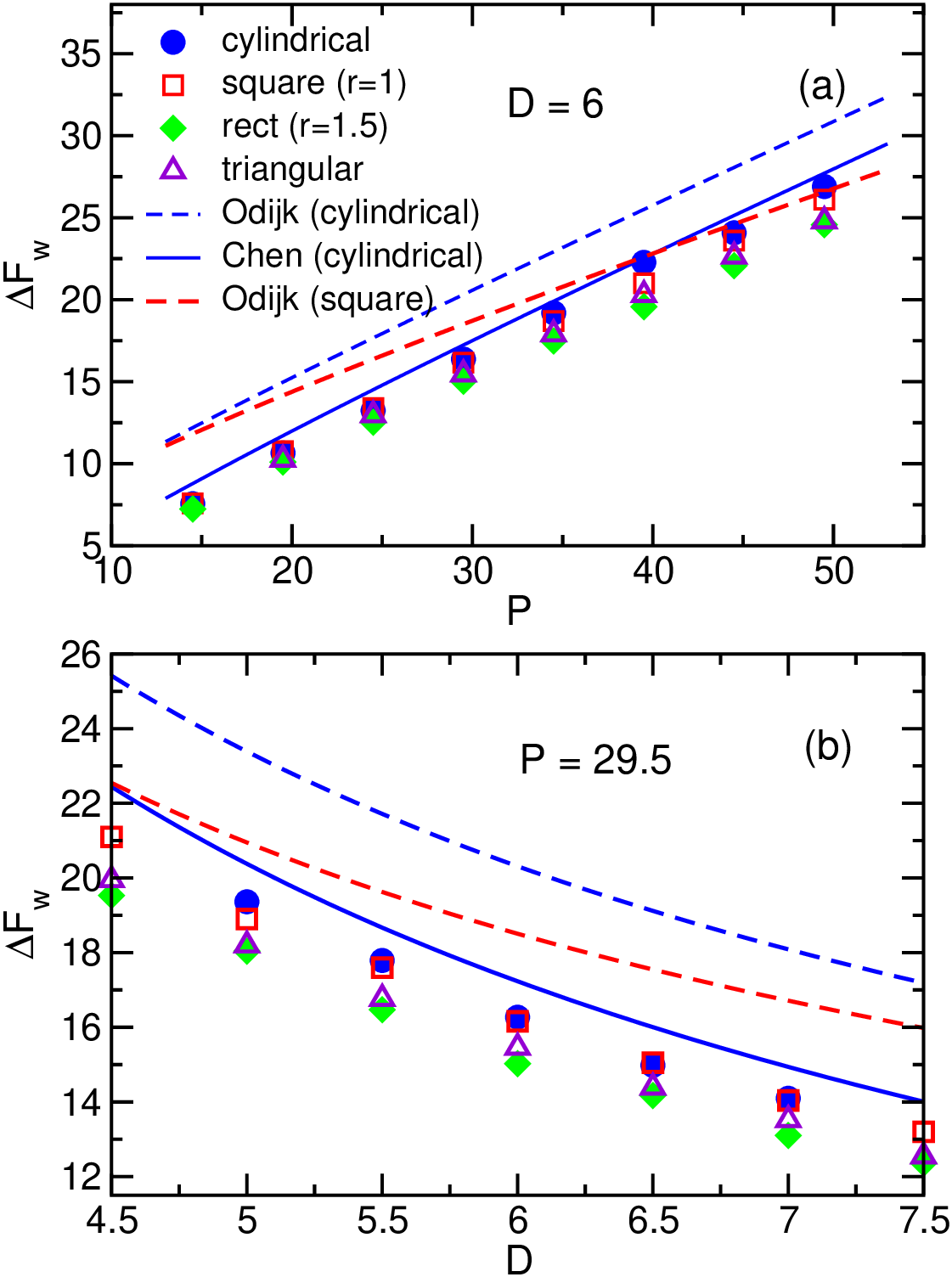}
\end{center}
\caption{(a) Hairpin free energy, $\Delta F_{\rm w}$, vs $P$ for a polymer in channels of
circular, square, rectangular and (equilateral) triangular cross-section.  Simulation results 
are shown for polymers of length $N$=200 and $D$=6.  Also shown are the predictions from the 
theory of Odijk\cite{odijk2006dna} for cylindrical and square channels, as well as the 
predictions of Chen\cite{chen2017conformational} for cylindrical channels. 
(b) As in (a) except $\Delta F_{\rm w}$ vs $D$ for polymers of persistence length $P$=29.5.  
The data for cylindrical channels are taken from Ref.~\citenum{polson2017free}.}
\label{fig:delFw}
\end{figure}

Figure~\ref{fig:delFw}(a) shows the variation of the free energy well depth $\Delta F_{\rm w}$
with $P$ for fixed $D$=6, and Fig.~\ref{fig:delFw}(b) shows the variation with respect to $D$
for fixed $P$=29.5. The well depth monotonically increases with increasing $P$  and decreases
monotonically with increasing $D$ for all channel geometries. At any given $D$ and $P$ the
values of $\Delta F_{\rm w}$ for different geometries are very close, though the triangular and
rectangular geometries generally have somewhat lower values. Finally, the inset of 
Fig.~\ref{fig:f.width} shows that $\Delta F_{\rm w}$ is independent of the polymer width.

As noted earlier, the well depth is a measure of the free energy cost for the formation of a 
hairpin turn. Two different theoretical studies thus far have examined the behavior
of this quantity. In one study, Odijk developed a mechanical model that considered
a hairpin turn whose plane is aligned with channel axis.\cite{odijk2006dna} Entropic depletion 
of the hairpin plane in relation to its transverse motion and azimuthal orientation was used to 
minimize the free energy and find the associated hairpin turn size for a given channel width 
and persistence length. Since the model neglects fluctuations in both the size and orientation
of the hairpin turn with respect to the channel axis, it is not expected to be of high 
accuracy, especially in the regime where $P\approx D$. This was borne out in the simulation 
results of Muralidhar and Dorfman in their study of the backfolded Odijk regime, where an 
indirect measurement of the hairpin free energy yielded a value that was approximately 
5~$k_{\rm B}T$ lower than the predicted value.\cite{muralidhar2014backfolding,%
muralidhar2016backfolded} More recently, Chen carried out a study using the Green's function 
formalism to determine the hairpin free energy and the global persistence length. Results obtained 
using the latter were found to be consistent with the simulation results of 
Ref.~\citenum{muralidhar2014backfolding}.

Theoretical predictions of Odijk's model for cylindrical and square channels, as well as 
a plot of an analytical representation to Chen's numerical solution for cylindrical channels are
overlaid on the data in Fig.~\ref{fig:delFw}. Both predictions capture the same general
trends, i.e. $\Delta F_{\rm w}$ increases with $P$ and decreases with $D$ in roughly the
same manner as the simulation data. However, the degree of quantitative agreement with the
simulation results differs. In the case of cylindrical tubes (as reported earlier in 
Ref.~\citenum{polson2017free}) the predictions of Chen are within 1 $k_{\rm B}T$ of
the measured results, while Odijk's theory predicts a hairpin free energy that is 
consistently overestimated by approximately 4$k_{\rm B}T$. In the case of square
channels, Odijk's predictions yield a comparable overestimate of the hairpin free energy,
except for very stiff chains. Note that Chen's study does not report results for square
or rectangular channels, and no predictions are available for triangular channels. 
It is noteworthy that Odijk's model yields values for cylindrical and square channels
that differ significantly from one another (except at low $P$), in contrast to the simulation 
results.  Finally, neither theory predicts any change in $\Delta F_{\rm w}$
with the chain width $w$, in agreement with the trend in the data shown in the inset 
of Fig.~\ref{fig:f.width}. This is to be expected, since $w$ is relevant only to 
excluded volume interactions, which are not significant in a hairpin turn.
The observed invariance of $\Delta F_{\rm w}$ with respect to $w$ also suggests that 
the discretization of the chain model used in the simulations (in contrast with the continuum 
polymer model used in the theoretical studies) is not a likely cause of the discrepancy.

\begin{figure}[!ht]
\begin{center}
\includegraphics[width=0.43\textwidth]{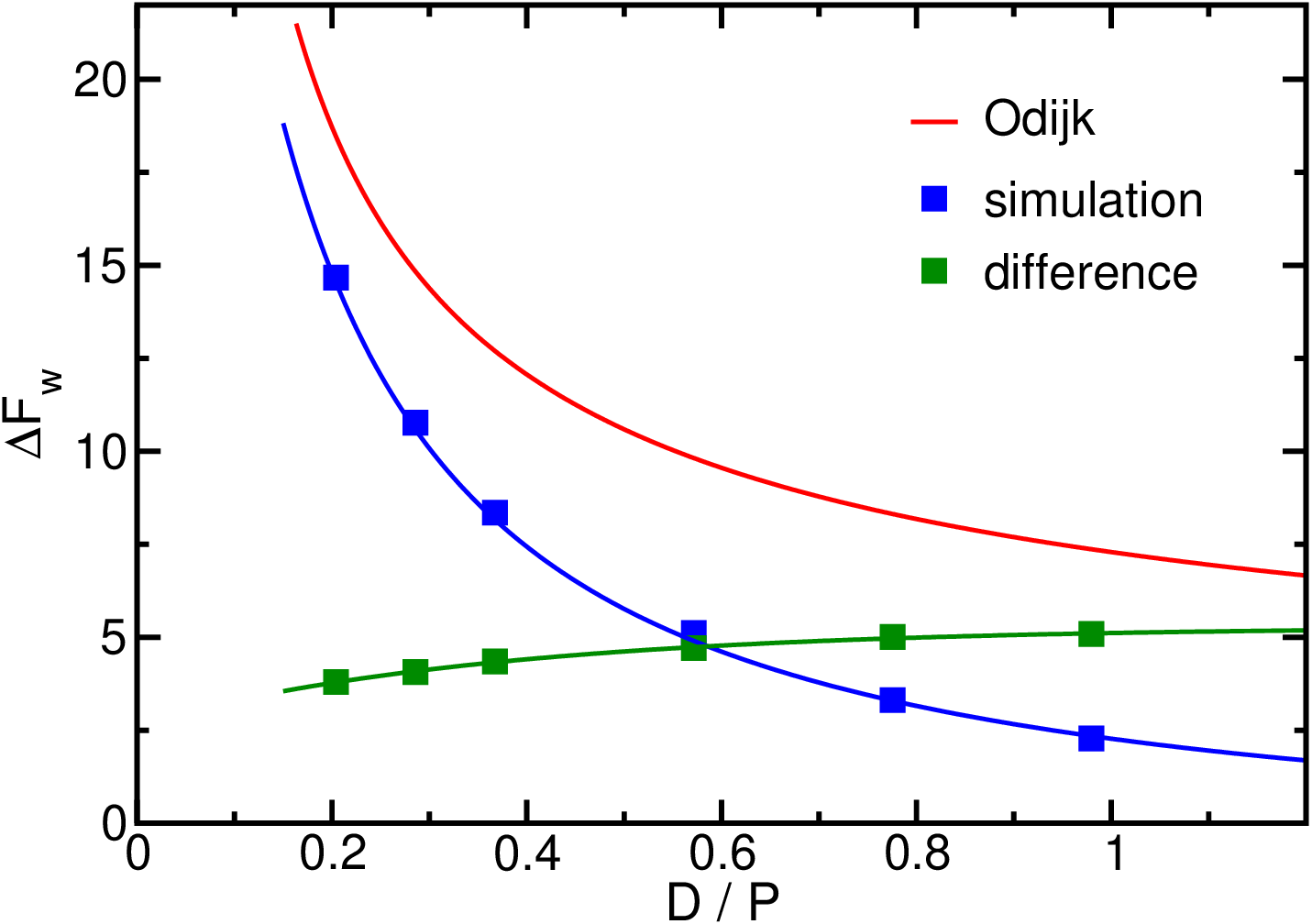}
\end{center}
\caption{Hairpin free energy vs $D/P$ for a polymer in a square channel. Simulation results
are shown for polymers of length $N$=100 and persistence length $P$=24.5. Also shown are
the predictions from the theory of Odijk in Ref.~\citenum{odijk2006dna} and the difference
between these predictions and the simulation results.
}
\label{fig:Fhp.square.N100}
\end{figure}

The simulation results presented above correspond to the case where the conditions required
for the Odijk regime, i.e. $P\gg D$, are marginally satisfied. However, there has recently been
considerable interest in the case where $P\approx D$, in which regime the backfolded
Odijk regimes are present.\cite{muralidhar2014backfolding,muralidhar2016backfolded,%
muralidhar2016backfolding} PERM simulations of a channel-confined semiflexible polymer in this 
regime were used to extract the global persistence length and thus the hairpin free energy.
As noted earlier, these calculations yield values that are approximately 5~$k_{\rm B}T$ lower
than Odijk's prediction.\cite{muralidhar2016backfolded} Figure~\ref{fig:Fhp.square.N100} shows
the variation of the hairpin free energy values for square channels with respect to the ratio $D/P$.
Also shown are Odijk's prediction and the difference between the theory and simulation.
We observe a difference of approximately 5~$k_{\rm B}T$ in the regime where $P\approx D$, in 
agreement with Ref.~\citenum{muralidhar2016backfolded}.

\begin{figure}[!ht]
\begin{center}
\includegraphics[width=0.43\textwidth]{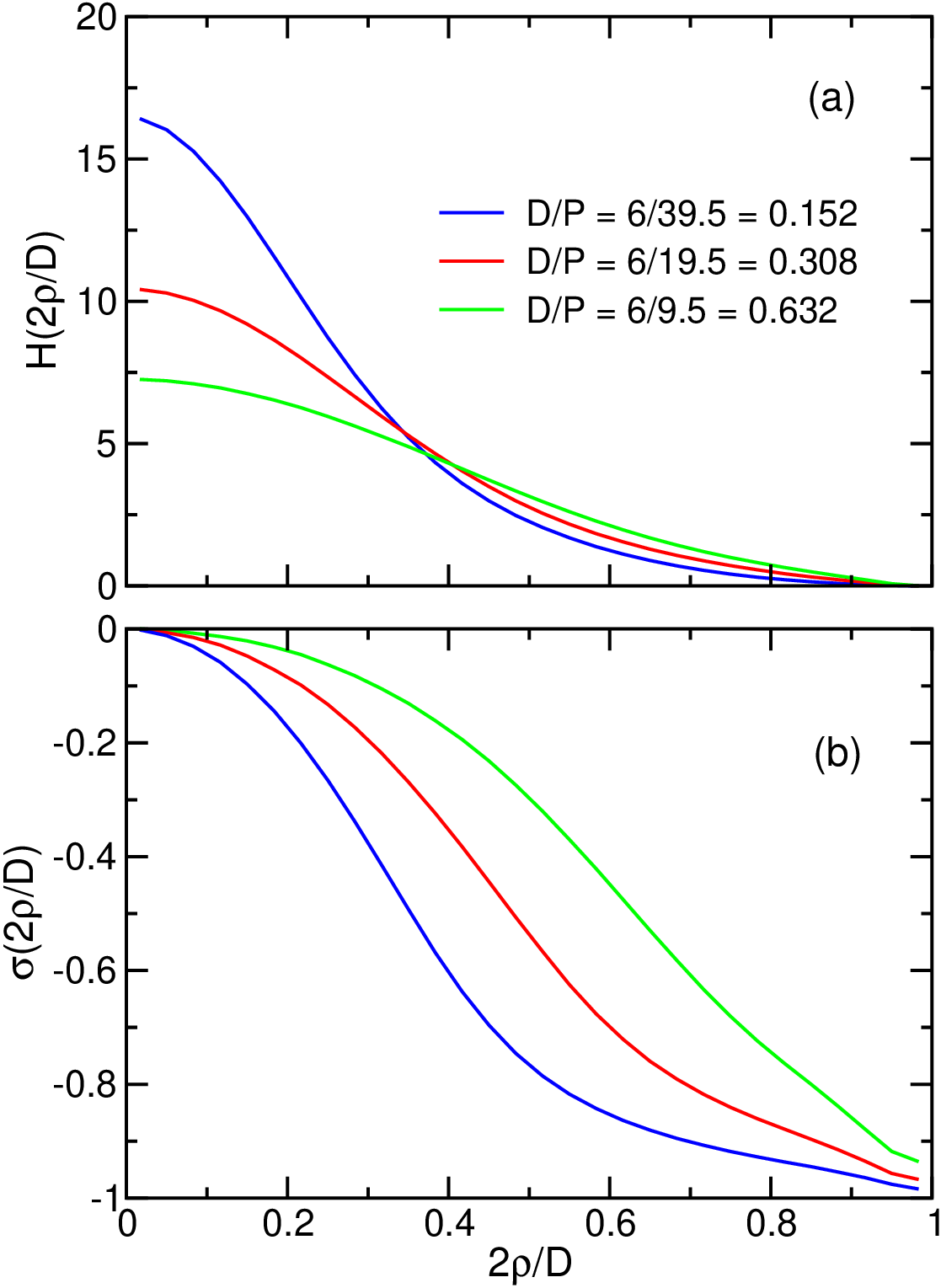}
\end{center}
\caption{(a) Hairpin-tip radial distributions for a  polymer of length $N$=100 in a 
cylindrical channel of diameter $D$=6. Results for various persistence lengths are shown.
(b) $\sigma$ vs scaled radial distance in a cylindrical channel, where $\sigma$ is a
parameter measuring the directional ordering of the tangent vector at the hairpin tip 
in the transverse plane, as defined in the text and in Ref.~\citenum{chen2017conformational}.
}
\label{fig:p.sig.rho.N100.D6}
\end{figure}

As noted in Ref.~\citenum{chen2017conformational}, the poorer predictions of Odijk's
theory in relation to those of Chen arises from the neglect in the former theory to explicitly
account for fluctuations in the hairpin size and the orientation of the hairpin plane 
with respect to channel axis.  Figure~\ref{fig:p.sig.rho.N100.D6}(a) shows the probability 
distributions of the position of the hairpin tip, $H(2\rho/D)$, for semiflexible polymers 
in a cylindrical tube. Here, $\rho$ is the transverse distance of the hairpin tip away 
from the central channel axis.  
Results are shown for different values of $D/P$.  In each case, the distribution is peaked 
at $\rho$=0, though there is an appreciable probability that the hairpin
tip is located away from the center of the channel. The distributions widen with decreasing 
relative chain stiffness. The results are in agreement with those reported by 
Chen.\cite{chen2017conformational} Figure~\ref{fig:p.sig.rho.N100.D6}(b) shows the variation
of the quantity $\sigma$ with $D/P$, where $\sigma\equiv \langle\cos2\phi\rangle$, and where
$\phi$ is the angle between the tangent vector ${\bf u}$ at the hairpin tip and the radial 
line segment connecting the tip to the central axis of the channel (see Fig.~4 of 
Ref.~\citenum{chen2017conformational}). This quantity is a measure of the degree of
anisotropy of the projections of ${\bf u}$ relative to the radial direction. For hairpins
with tips close to the central axis the projections are isotropic and $\sigma$=0,
while strong directional alignment is observed when the tip is close to the wall, where
$\phi$ tends to $\pi/2$ and $\sigma\rightarrow 1$. The curves show a transition between
these two limits. As the polymer stiffens and $D/P$ decreases, the entire hairpin becomes
more coplanar while maximizing the hairpin size. Hairpins whose tips lie close to the wall
become rarer but even more ordered, leading to $\sigma$ decrease closer to the limit of $-1$. 
The result is the qualitative trend of a reduction of $\sigma$ with increasing $D/P$, while 
maintaining the boundary conditions of $\sigma=0$ at $\rho$=0 and $\sigma=-1$ at $\rho$=$D/2$.
The curves are consistent with those presented in Ref.~\citenum{chen2017conformational}.

\begin{figure}[!ht]
\begin{center}
\includegraphics[width=0.46\textwidth]{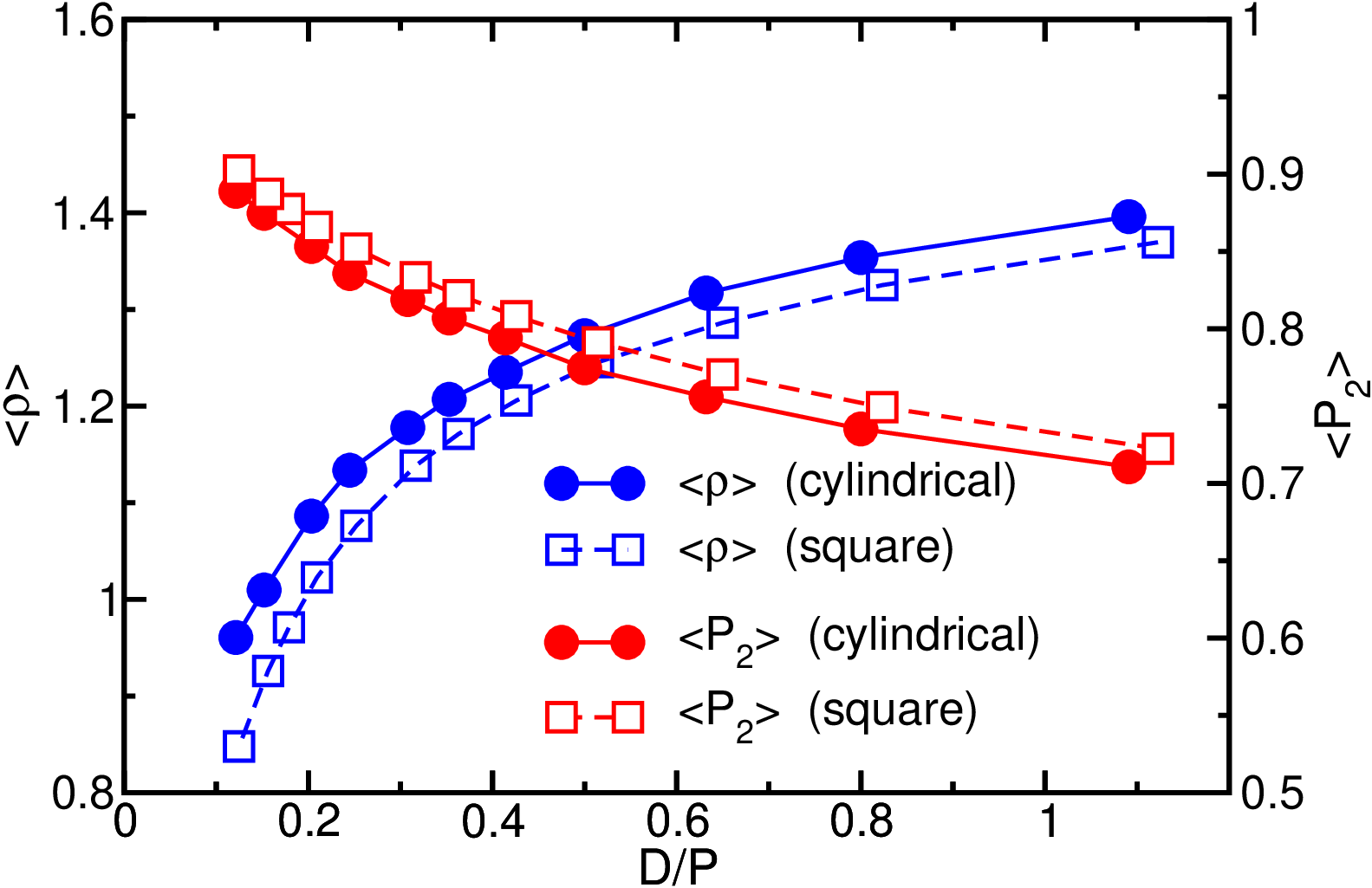}
\end{center}
\caption{Hairpin orientational order parameter $\langle P_2\rangle$ and mean radial
hairpin distance $\langle\rho\rangle$ vs $D/P$. Results are shown for $N$=100 and $D$=6
for cylindrical and square nanochannels.
}
\label{fig:p2.rho.av.N100.D6}
\end{figure}

Figure~\ref{fig:p2.rho.av.N100.D6} shows the mean radial displacement of the tip,
$\langle\rho\rangle$, as a function of $D/P$. Overlaid on the same graph is the orientational
order parameter of the hairpin plain at the location of the tip, $\langle P_2\rangle \equiv
\langle P_2(\cos\theta)\rangle$, where $\theta$ is the angle of the plane relative to the
channel axis. Results are shown for both cylindrical and square channels. The increase
in $\langle\rho\rangle$ with $D/P$ reflects the narrowing of the distributions with $D/P$
observed in the case of cylindrical channels in Fig.~\ref{fig:p.sig.rho.N100.D6}. The
order parameter $\langle P_2\rangle$ decreases as the chains stiffen. Note that 
over the entire range of $D/P$ considered here, $\langle P_2\rangle$ is appreciably
lower than the value $\langle P_2\rangle=1$, the value corresponding to perfect alignment
of the plane with the channel assumed in Odijk's model. As noted by 
Chen\cite{chen2017conformational}, the neglect of the hairpin plane orientational fluctuations 
that lead to $\langle P_2\rangle<1$ is the main factor in the underestimate of the hairpin 
entropy and the corresponding overestimate of the free energy.

\begin{figure}[!ht]
\begin{center}
\includegraphics[width=0.43\textwidth]{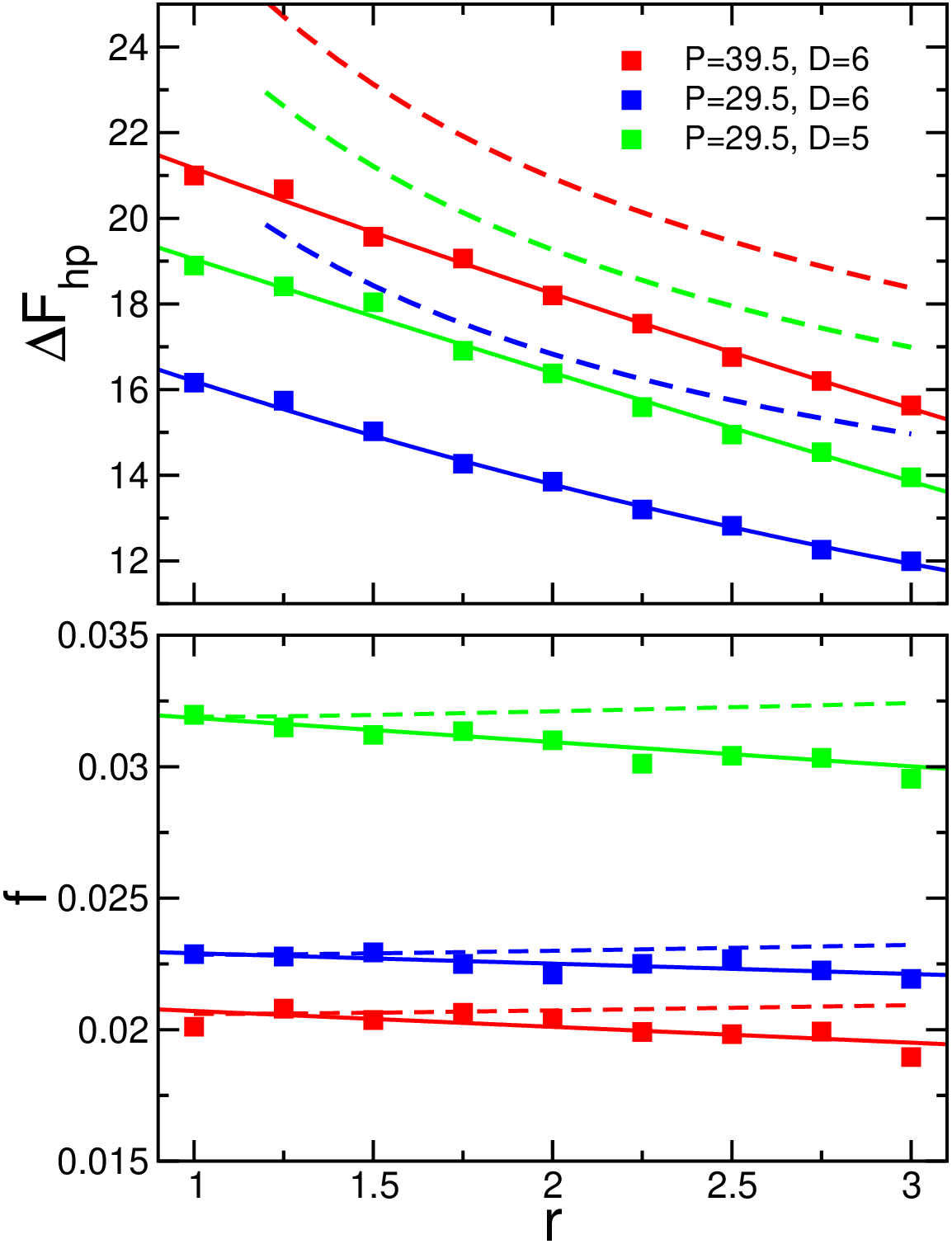}
\end{center}
\caption{(a) Hairpin free energy, $\Delta F_{\rm w}$, vs aspect ratio $r$ of a rectangular
nanochannel.  Results are shown for a polymer with $N$=200 in channels of different 
cross-sectional areas and for polymers of different $P$.  The solid lines of the same 
color are guides for the eye, while the dashed lines of the same color are predictions 
from Odijk.\cite{odijk2008scaling} (b) The entropic force $f$ vs channel aspect ratio $r$.
The solid lines are guides for the eye. The dashed lines are predictions 
calculated using an approximation that employs Eq.~(\ref{eq:B3}) to estimate the 
scaling of the angle $\delta$ between interacting deflection segments in the calculation
of the excluded volume.} 
\label{fig:delFw.f.ratio}
\end{figure}

Thus far, most of the results presented have been for maximally symmetric cases
of a given channel geometry, i.e channels with circular, square and equilateral triangular 
cross sections. Given the importance of asymmetric geometries employed in experiments for
both rectangular\cite{werner2018hairpins} and triangular\cite{%
huh2007tuneable,fanzio2011dna,angeli2011dna,fanzio2012modulating,manneschi2013conformations,%
manneschi2014stretching} channels it is of interest to examine the effects of such 
asymmetry on the free energy gradient and hairpin free energy.
In Figs.~\ref{fig:dFdk.P.D} and \ref{fig:delFw} it was shown that $f$ and $\Delta F_{\rm w}$
for polymers in rectangular channels with an asymmetric ratio of $r$=1.5 scale with $D$ 
and $P$ in manner consistent with that for square channels. Figure~\ref{fig:delFw.f.ratio}(a)
shows the variation of $\Delta F_{\rm w}$ with respect to $r$, while 
Fig.~\ref{fig:delFw.f.ratio}(b) shows the corresponding variation of $f$. 
Results are shown for two different values of $P$ and two values of $D$. 
In each case, $\Delta F_{\rm w}$ decreases approximately linearly with 
increasing asymmetry, while $f$ is almost constant, though it does decrease very 
slightly with $r$. The dashed curves overlaid on the data in Fig.~\ref{fig:delFw.f.ratio}(a)
show the predictions for the hairpin free energy using Odijk's model for rectangular 
channels.\cite{odijk2006dna} Generally, the predictions are qualitatively accurate in
predicting a monotonic decrease in $\Delta F_{\rm w}$ with $r$, though they consistently 
overestimate the value by 2--3 $k_{\rm B}T$ over the range examined. Note that the
theoretical predictions for rectangular channels use the approximation of very large
asymmetry in the calculation of the entropic contribution to the free energy.\cite{odijk2006dna}
Consequently, the results cannot be extrapolated to $r$=1. 

How can we understand the fact that $f$ is nearly constant with respect to $r$ at fixed
$D$ and $P$? Recall that Eq.~(\ref{eq:fscale}) was derived for 
symmetric channels characterized by a single lateral dimension.  However, for asymmetric 
rectangular channels there are two independent channel widths, $D_{\rm W}$ and $D_{\rm H}$, 
which define the asymmetry, $r\equiv D_{\rm W}/D_{\rm H}\geq 1$.
To determine the scaling for $f$ with respect to $r$, we first consider the limiting case 
where $D_{\rm W}\gg D_{\rm H}$. As noted by Odijk,\cite{odijk2008scaling} 
the angle $\delta$ describing the relative orientation between interacting
deflection segments satisfies $\langle|\sin\delta|\rangle \sim (D_{\rm W}/l_{\rm d})^2$ 
for this case. Otherwise, the arguments used to derive Eq.~(\ref{eq:fscale}) are also 
applicable for rectangular channels. Noting that the cross-sectional area is 
$A = D_{\rm W}D_{\rm H} \propto D^2$, it can easily be shown that:
\begin{eqnarray}
f \sim w D^{-5/3}P^{-1/3} r^{1/6}
\label{eq:fscale2}
\end{eqnarray}
Thus, $f$ is predicted to scale with $P$, $w$ and $D$ (for fixed asymmetry $r$) in the same 
manner as for symmetric channels. As shown in Fig.~\ref{fig:dFdk.P.D}, the measured scaling for 
$P$ and $D$ for the case of rectangular channels with $r$=1.5 is close to this prediction.
However, Eq.~(\ref{eq:fscale2}) also predicts that $f$ monotonically increases with $r$ by 
$\approx$20\% over the range $r=1-3$ considered. This constitutes a significant quantitative 
discrepancy with respect to the simulation result and no doubt arises from imposing the
condition $D_{\rm W}/D_{\rm H}\gg 1$ in the derivation.  To derive a better prediction, 
we require a more accurate relation for the dependence of the angle $\delta$ 
on the channel dimensions and persistence length that is valid for smaller values of
$D_{\rm W}/D_{\rm H}$.  This provided by Eq.~(B3) of Ref.~\citenum{odijk2008scaling}:
\begin{eqnarray}
\langle|\sin\delta|\rangle \approx \sqrt{\frac{G_{\rm D}+G_{\rm A}+1}{(G_{\rm D}+1)(G_{\rm A}+1)}},
\label{eq:B3}
\end{eqnarray}
where $G_{\rm A}\equiv P/D_{\rm W}$ and $G_{\rm D}\equiv P/D_{\rm H}$. As is evident from 
Fig.~\ref{fig:delFw.f.ratio}(b), the predicted free energy gradient is nearly constant with 
respect to variation in the asymmetry over the range $r=1-3$. While the theoretical prediction 
fails to reproduce the observed decrease in $f$ with increasing $r$, it is much more accurate than 
that obtained using Eq.~(\ref{eq:fscale2}), as expected. The remaining discrepancy is
likely due to remaining inadequacies in the theoretical model, such as treating the deflection 
segments as rigid rods and describing the interactions between them using the 2nd virial 
approximation.

\begin{figure}[!ht]
\begin{center}
\includegraphics[width=0.43\textwidth]{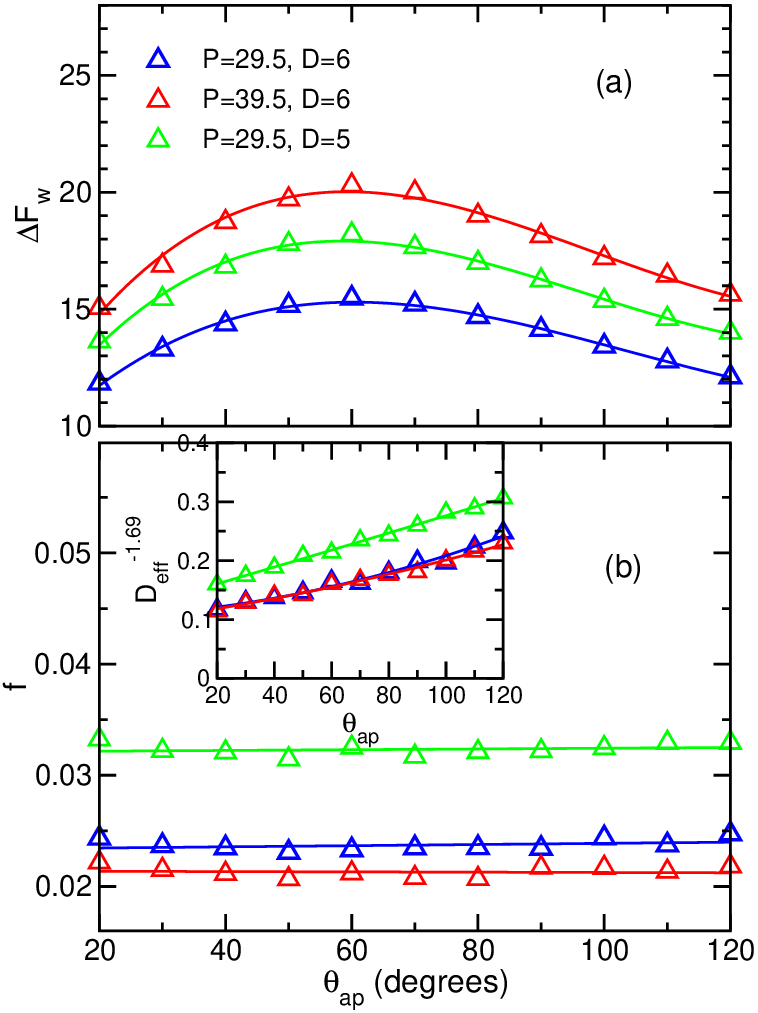}
\end{center}
\caption{(a) Hairpin free energy, $\Delta F_{\rm w}$, vs apex angle $\theta_{\rm ap}$ for 
a polymer in a triangular nanochannel. Results are shown for a polymer with $N$=200 in channels
of different cross-sectional areas and for polymers of different $P$. The solid lines
of the same color are guides for the eye. (b) The entropic force $f$ vs apex angle 
$\theta_{\rm ap}$ obtained from the same free energy functions used in (a). The solid lines 
are guides to the eye.  The inset shows $D_{\rm eff}^{1.69}$  vs $\theta_{\rm ap}$, where 
the effective channel dimension $D_{\rm eff}$ is defined in the text.}
\label{fig:delFw.f.tri_angle}
\end{figure}

Let us now consider the effects of asymmetry of triangular channels. Specifically,
we examine channels whose cross sections are isosceles triangles with an apex
angle of $\theta_{\rm ap}$ and the two other angles of $(\pi-\theta_{\rm ap})/2$ each. 
Figure~\ref{fig:delFw.f.tri_angle}(a) and (b) show the variation of $\Delta F_{\rm w}$
and $f$ with respect to $\theta_{\rm ap}$, respectively. Results are shown for different values of
$P$ and $D$. Again we note the general result that $\Delta F_{\rm w}$ increases
and $f$ decreases with $D$ and $P$ for all values of $\theta_{\rm ap}$, again consistent
with the results for symmetric channels in Figs.~\ref{fig:dFdk.P.D} and \ref{fig:delFw}.

In each case, the hairpin free energy peaks at $\theta_{\rm ap}=60^\circ$. This is consistent 
with the expectation that the hairpin plane aligns roughly along the wall between 
two of the corners of the triangle. For a triangle with different side lengths,
the plane is expected to align along the channel wall(s) with the largest side length
in order to maximize the hairpin size and thus minimize the energy. 
When $\theta_{\rm ap}$ is varied for fixed cross-sectional area, the minimum value of 
this distance between any two corners occurs for an equilateral triangle, i.e.
all angles of $60^\circ$. As $\theta_{\rm ap}$ is increased or decreased from this value,
the maximum side length of a channel wall increases, and so the energy decreases.
Obviously, the variation in the entropy associated with the azimuthal and 
longitudinal orientational freedom also contributes to $\Delta F_{\rm w}$, but 
the hairpin energy appears to be the dominant factor affecting the qualitative
trend for $\Delta F_{\rm w}$.

Figure~\ref{fig:delFw.f.tri_angle}(b) shows that the free energy gradient is independent
of the channel asymmetry.  This is a somewhat surprising result for triangular channels, where 
the effects of entropic depletion near the corners have been shown previously to be 
appreciable relative to the case of rectangular channels.\cite{reinhart2013entropic,%
manneschi2013conformations} As noted earlier, entropic depletion leads to a reduction 
in the effective cross-sectional area of the channel, 
which in turn should effectively increase the channel dimension $D$. Reinhart {\it et al.}
developed a measure of the effective area that uses the monomer probability distribution
in the transverse plane of the channel.\cite{reinhart2013entropic} They took the 
effective channel width to be the geometric average of the full width at 
half-maximum of the projections of the distributions onto the two axes defined by the
eigenvectors of the 2D distribution. We have calculated the same quantity, which we
call $D_{\rm eff}$, as a function of apex angle $\theta_{\rm ap}$ for channels of a given 
dimension $D$ and for polymers of a given $P$. Naively using $D_{\rm eff}$
in place of $D$ in Eq.~(\ref{eq:fscale}) or (\ref{eq:fscale2}), 
one expects $f\sim D_{\rm eff}^{-5/3}$,
The inset of Fig.~\ref{fig:delFw.f.tri_angle} shows $D_{\rm eff}^{-1.69}$
vs $\theta_{\rm ap}$, where the exponent value is chosen to match the scaling of $f$ with respect
to $D$ in Fig.~\ref{fig:dFdk.P.D}(b) for equilateral triangular channels (using an
exponent of 5/3 does not appreciably change the result). For each case of $P$ and $D$ 
shown, the quantity $D_{\rm eff}^{-1.69}$ increases with $\theta_{\rm ap}$ roughly by a factor 2
over the range of apex angles considered. This suggests that $f$ should increase by roughly 
this factor as well, in contrast to the measured invariance of $f$ with respect to $r$. 
Thus, this approach fails to account for the trends observed in the present case, 
in contrast with the case in Ref.~\citenum{reinhart2013entropic}. This may arise
from the fact that we consider much larger $P/D$ ratios in the present study 
that were used in Ref.~\citenum{reinhart2013entropic}.

\begin{figure}[!ht]
\begin{center}
\includegraphics[width=0.43\textwidth]{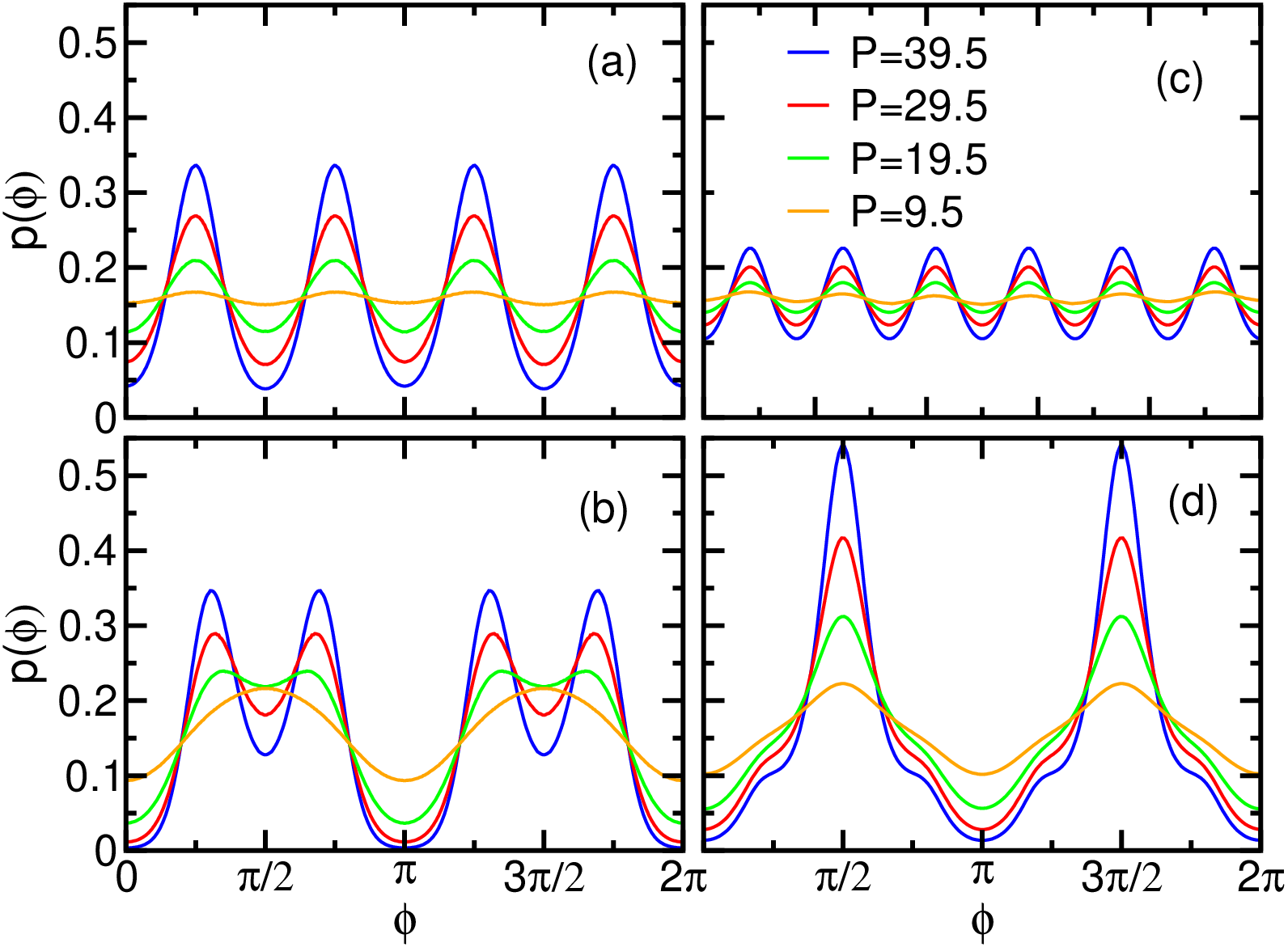}
\end{center}
\caption{ Distribution $p(\phi)$ of the azimuthal angle $\phi$ for the orientation of
the hairpin plane in the plane transverse to the square channel.  Results are shown for a
polymers of length $N$=100 in channels of area $D$=6 for various persistence lengths.
Results in the separate graphs are for the following cases: (a) square channel; (b) rectangular 
channel with $r$=1.5; (c) equilateral triangular channel (i.e. $\theta_{\rm ap}$ = $60^\circ$); 
and (d) triangular channel with apex angle of $\theta_{\rm ap}$=70$^\circ$.}
\label{fig:pphi.N100.D6}
\end{figure}

Finally, it is of interest to investigate the effects of channel asymmetry on the
orientational behavior of the hairpin plane, which is an important part of the
entropic contribution to $\Delta F_{\rm w}$.  Figure~\ref{fig:pphi.N100.D6} shows
the probability distribution $p(\phi)$ for (a) square, (b) rectangular, (c) equilateral
triangular, and (d) asymmetric triangular channels. Here, $\phi$ is the azimuthal angle
describing the hairpin plane orientation around the channel axis. All the results
are shown for a fixed channel dimension of $D$=6. In each case distributions are
shown for four different persistence lengths. As expected, Fig.~\ref{fig:pphi.N100.D6}(a)
shows that there are four preferred orientations of the hairpin plane, each located along
the diagonal of the square, where the hairpin size is largest (i.e. there are two diagonal
directions, each of which corresponds to two hairpin orientations 180$^\circ$ apart). 
The degree of alignment of the plane along the diagonal decreases as the polymer becomes
more flexible and the distribution becomes more uniform. The distributions for asymmetric
(i.e. rectangular) channels in Fig.~\ref{fig:pphi.N100.D6}(b) are qualitatively similar
to those for the square channels for stiff polymers. As expected, however, the peaks
shift to different orientations that correspond to the diagonals of the rectangle. 
More interestingly, for sufficiently flexible polymers (i.e. when $P\approx D$) the
preferred orientation becomes parallel to the long side of the rectangle. This orientation
likely has a slightly higher energy than for an orientation along the diagonal. However,
the energy contributes less to the free energy for more flexible chains, and its increase is
likely compensated for by an increase in the entropy from the longitudinal orientational
fluctuations of the hairpin plane. 

Figure \ref{fig:pphi.N100.D6}(c) shows the distributions for equilateral triangular
channels. The six peaks arise at orientations corresponding to the alignment of
the plane close to and along a channel wall between two of the corners (i.e. three
identical walls, each associated with two hairpin plane orientations related by a 
$180^\circ$ rotation).
As for the square channels, the preference for this alignment weakens as the polymers
become more flexible and, thus, the distributions become more uniform. 
Figure~\ref{fig:pphi.N100.D6}(d) shows distributions for an isosceles triangle 
with an apex angle of $\theta_{\rm ap}=70^\circ$. For such a seemingly small asymmetry,
the effects are pronounced. The preferred orientation are at the two angles
where the hairpin plane is parallel to the long side of the triangle, i.e.
the side opposite to the apex. The slight preference for orientations parallel
to the other walls is evident only in the weak shoulders of the main peaks.
Overall, the contrast between the sharp peaks in Fig.~\ref{fig:pphi.N100.D6}(d) 
and the more uniform distributions of if Fig.~\ref{fig:pphi.N100.D6}(c) suggest
a considerable decrease in orientational entropy with increasing asymmetry. The
fact that $\Delta F_{\rm w}$ peaks at $\theta_{\rm ap}=60^\circ$ suggests that this is more
than compensated for by the reduction in energy that comes with larger hairpins. 

\section{Conclusions}
\label{sec:conclusions}

In this study, we have used Monte Carlo simulations to characterize the conformational
free energy of folded polymers confined to long nanochannels. By measuring the free energy
$F$ of a polymer as a function of the end-to-end distance $\lambda$, we determined
the free energy gradient $f\equiv|dF/d\lambda|$ for polymers in states of internal
overlap, as well as the free energy of the hairpin fold. In addition to similar
calculations carried out in our previous study using cylindrical channels\cite{polson2017free} 
we have studied confinement effects for rectangular and triangular channels as well. 
Generally, we find that $f$ scales with the channel width $D$, persistence length $P$ and 
effective polymer width $w$ in a manner consistent with the predictions of Eq.~(\ref{eq:fscale}).
Small deviations in the scaling exponents likely arise from the fact that the conditions for 
various approximations are only marginally satisfied. Notably, the scaling appears to be
independent of the type of channel geometry (i.e. cylindrical, rectangular or triangular). 
In addition, $f$ is not appreciably affected by varying the asymmetry of rectangular or 
triangular channels. The hairpin free energy was not strongly affected by the channel
geometry type, though it does exhibit a decrease with increasing channel asymmetry for
both rectangular and triangular channels. The scaling of the hairpin free energy with
$P$, $D$ and channel asymmetry $r$ for rectangular channels was qualitatively consistent with 
predictions from Odijk's theory,\cite{odijk2006dna,odijk2008scaling} though the quantitative
discrepancy was significant. The origin of this discrepancy is the neglect of
hairpin shape and orientational fluctuations, as elucidated by Chen\cite{chen2017conformational} 
for cylindrical channels and corroborated by analysis of hairpin behavior in this study.
Finally, the difference between the predicted and measured hairpin free energy of 
$\approx$5~$k_{\rm B}T$ from simulations for polymers the backfolded Odijk 
regime\cite{muralidhar2014backfolding,muralidhar2016backfolded} is consistent with 
the value measured in the present work.

The quantitative characterization of the free energy functions for nanochannel-confined
polymers is useful for validating the accuracy of theoretical models used to analyze
and interpret experimental results of confined DNA. For example, the expression for the 
free energy gradient in Eq.~(\ref{eq:fscale}) has been employed recently in the analysis of 
both the equilibrium fluctuations in the extension length for single-fold
DNA\cite{werner2018hairpins} in rectangular channels and the nonequilibrium unfolding 
dynamics in square nanochannels.\cite{reifenberger2018odijk} 
It is also of value in providing independent confirmation of findings of other simulation 
studies, such as the observed discrepancy between the predicted and measured hairpin
free energy in simulations of the backfolded Odijk regime for rectangular channels.
To our knowledge, this is the first direct study of folding properties of polymers in 
triangular channels, a geometry which has been used extensively in DNA extension experiments.
In future work, it will be of value to further examine folding in nanochannels by 
employing a model that better accounts for the effects of electrostatic 
interactions present in DNA experiments on the polymer-channel interactions, such as
those used in Refs.~\citenum{cheong2017wall} and \citenum{manneschi2014stretching}.

\begin{acknowledgement}
This work was supported by the Natural Sciences and Engineering Research Council of Canada 
(NSERC).  We are grateful to the Atlantic Computational Excellence Network (ACEnet), WestGrid
and Compute Canada for use of their computational resources.
\end{acknowledgement}

\providecommand{\latin}[1]{#1}
\makeatletter
\providecommand{\doi}
  {\begingroup\let\do\@makeother\dospecials
  \catcode`\{=1 \catcode`\}=2 \doi@aux}
\providecommand{\doi@aux}[1]{\endgroup\texttt{#1}}
\makeatother
\providecommand*\mcitethebibliography{\thebibliography}
\csname @ifundefined\endcsname{endmcitethebibliography}
  {\let\endmcitethebibliography\endthebibliography}{}


\end{document}